\documentclass[12pt,epsfig]{article}

\usepackage{amssymb,epsfig}

\newcommand{\be}{\begin{equation}}

\newcommand{\ee}{\end{equation}}

\newcommand{\bea}{\begin{eqnarray}}

\newcommand{\eea}{\end{eqnarray}}

\begin{document}

\begin{center}

{\LARGE \bf   Generalized Parton Distributions and Generalized Distribution
Amplitudes : \\[1mm] New Tools for Hadronic
Physics\footnote{Talk given at the International Conference
on Theoretical Physics, TH2002, Paris, July 2002. }}
\vspace{1cm}

{\sc B.~Pire},
\\[0.5cm]
\vspace*{0.1cm}
{\it
CPhT, {\'E}cole Polytechnique, F-91128 Palaiseau, France\footnote{
  Unit{\'e} mixte C7644 du CNRS.}
                       } \\[1.0cm]

\vskip2cm
\end{center}

\begin{abstract}
{\small The generalized parton distributions and the
generalized distribution amplitudes give access to a deeper understanding of
the quark and gluon content of hadrons. In this short review, we select
some new
developments of their interesting connections with the physics information that
one can extract from exclusive reactions  at medium
and high energies. }
\end{abstract}

\vskip1cm

\section{Generalized Parton Distributions}

A considerable amount of theoretical and experimental work is
currently being devoted to the study of generalized parton
distributions, which are defined as Fourier transforms of matrix elements
between different hadron states, such as:
$$
\int d\lambda e^{i\lambda x(p+p').n} \langle N'(p',s') |\bar\psi(-\lambda
n)\gamma.n
\psi(\lambda n) | N(p,s) \rangle
$$

Their measurements are expected to yield important contributions
to our understanding of how quarks and gluons assemble themselves into
hadrons~\cite{GPD}. The
simplest and cleanest  exclusive processes where
these distributions occur are deeply virtual Compton scattering (DVCS),
i.e., $\gamma^* N \to \gamma N'$ and meson electroproduction i.e., $\gamma^* N
\to M N'$ in kinematics where the
$\gamma^*$ has large spacelike virtuality $Q^2$ while the invariant momentum
transfer
$t$ to the proton is small (for recent reviews, see \cite{rev} ).
Their ``inverse'' processes, $\gamma N \to \gamma^* N'$ and $\pi N \to \gamma^*
N'$ at small $t$ and large
\emph{timelike} virtuality of the final state photon~\cite{TCS} are quite
similar.

DVCS combines
features of the inelastic processes with those of an elastic process.
A relativistic charged lepton  is
scattered from a target nucleon or nucleus.  A real photon of
4-momentum $q_{\mu}'$ is also observed in the final
state.  With $e(k),\, e'(k')$ denoting the initial and final electrons of
momenta $k, \, k'$ respectively, and $p, \, p'$ denoting the momentum of the
target, the process is $$e(k)+ N(p) \to e'(k')+N'(p') +\gamma(q').$$ The net
momentum transfer $\Delta$ to the target is obtained by momentum
conservation, $\Delta ^{\mu} = k ^{\mu} -k^{' \, \mu } -q^{' \, \mu} $.
The real photon may be emitted by the lepton beam, in which case a
virtual photon of momentum $Q_{BH}^{\mu}=\Delta^{\mu}$ strikes the
target.  This is the Bethe-Heitler amplitude which is perfectly under
control since the nucleon form factors at small $t$ are
known.
A second possibility is that the target absorbs a virtual photon of
momentum $Q_{VCS}^{\mu}=p'^{\mu}-p^{\mu}+q^{'\,\mu }$ and emits the real
photon.
This is the genuine DVCS amplitude. It is straightforward to select events
where all components of
$\Delta ^{\mu}$ are small compared to $\sqrt{Q^{2}}$, with
$Q^{2}=-Q_{VCS}^{\mu}Q_{VCS, \, \mu} > GeV^{2}$.  These conditions
have recently been realized in experiments\cite{HermesCEBAF} at HERMES
at DESY and CEBAF at JLab.

The physical interpretation is that the target is resolved by the
virtual photon on a spatial scale small compared to the target size.
A photon of high virtuality $Q^{2}$ selects a short-distance region of
the target: the spatial resolution is of order $\Delta b_{T} \sim
\hbar/ \sqrt{ Q^{2}}$.
Perturbative QCD (pQCD) can be applied to DVCS at large $Q^{2}$ ,
exploiting the short-distance resolution of the virtual photon,
despite the presence of a real photon in the reaction\cite{GPD}.

Extracting the GPD's from the scattering amplitudes require to study a number
of observables in different reactions\cite{rev}; electroproduction of
mesons and
photons is the main source of information, but for the chiral-odd GPD's which
need a quite different process\cite{IPST}. GPD's may also be studied for the
deuteron\cite{deut} and other nuclei.

\section{Femtophotography}

Complementarily to ordinary parton distributions which measure the
probability  that a quark or gluon carry a fraction $x$ of the hadron momentum,
GPD's represent the interference of different wave functions, one where a
parton carries momentum fraction $x+\xi$ and one where this fraction is
$x-\xi$. $\xi$ is called the skewedness and is fixed in a DVCS experiment by
external momenta. On the contrary, $x$  is an integration variable varying from
$-1$ to $+1$. When $x < \xi$, the GPD's should be interpreted as the
interference of the hadron wave function with the wave function of the hadron
accompanied by a $q \bar q$ pair. It is thus very reminiscent of the
probability
amplitude to extract a meson from a hadron.

Apart from longitudinal momentum fraction variables, GPD's also depend on the
momentum transfer $t$ between the initial and final hadrons. Fourier
transforming this transverse momentum  information leads to information on the
transverse location of quarks and gluons in the hadron\cite{femto}. Real-space
images of the target can thus be obtained, which is completely new.  Spatial
resolution is determined by the virtuality of the incoming photon .  Quantum
photographs of the proton, nuclei, and other elementary particles with
resolution on the scale of a fraction of a femtometer are thus feasible.

\section{Generalized distribution amplitudes}
The crossed version of GPD's describe the exclusive hadronization of a $q \bar
q$ or $g g$ pair in a pair of hadrons, a pair of $\pi $ mesons for instance.
These generalized distribution amplitudes (GDA) \cite{DGPT}, defined in the
quark-antiquark case, as

$$
\Phi_q^{\pi\pi} (z,\zeta,s) =\int\frac{dx^-}{2\pi} e^{-izP^+x^-}
\langle \pi(p') \pi(p) |\bar\psi_q(x^-)\gamma^+
\psi_q(0) | 0) \rangle
$$
where $s$ is the squared energy of the $\pi \pi$ system, are the non
perturbative part of the light cone dominated process\cite{DGP}

$$
\gamma^* \gamma \to \pi \pi
$$
which may be measured in electron positron colliders of high luminosity.
This new QCD object allows to treat in a consistent way
the final state interactions of the meson pair. Its phase is related to the
the phase of $\pi \pi $ scattering amplitude, and thus contains information on
the resonances which may decay in this channel. Results on the related
reaction $\gamma^* \gamma \to \rho \rho$ may be
expected from LEP 2.

\section {Hunting for the Odderon}
A nice application of the GDA's concerns the search for the Odderon\cite{HPST}.
Pomeron and Odderon  exchanges are the theoretically dominant contributions to
hadronic cross sections at high energy. They appear on an equal footing in the
QCD description of hadronic reactions, and in the lowest order approximation
they correspond to  colour singlet exchanges in the $t$-channel with
 two and three gluons, respectively.
The Odderon remains a mistery from an experimental point of view.
On the one hand, recent studies of the elastic $pp$ scattering
show that one needs the Odderon contribution to understand the
data in the dip region \cite{Doshrecent}.
On the other hand, the studies of meson production processes which should
select the odderon exchange didn't show any clear signal of its
importance\cite{Olsson,Dosh}. In these cases, the
scattering amplitude describing Odderon exchange enters quadratically in the
cross section.

A number of interesting features of the two pion diffractive
electroproduction process allows to search for the QCD-Odderon at the amplitude
level.
Since the two pion state described by the GDA's doesn't have any definite
charge
parity, both Pomeron and Odderon exchanges contribute. The charge asymmetry is
ideally suited to select the interference of the two amplitudes. As in  open
charm production\cite{Brodsky},the Odderon amplitude enters linearly in the
asymmetries and therefore one can hope  that Odderon effects can show up  more
easily. Moreover factorization properties allow to perturbatively
calculate the short-distance part of the scattering amplitude. The long
distance part contains the product of the C-even and C-odd GDA's, and shows a
dramatic  dependence in the two pion invariant mass.

\vspace{1.5cm}

In conclusion, let me acknowledge that I would not have been able to give this
mini review without all the fruitful discussions I had during the last 6 years
with all my collaborators and particularly Markus Diehl, Thierry Gousset, John
Ralston, Lech Szymanowski and Oleg Teryaev.

\end{document}